\documentclass{aa}  

\usepackage{graphicx}
\usepackage{txfonts}
\begin{document}

   \title{A giant umbrella-like stellar stream around the tidal ring galaxy NGC~922}


   \author{David Mart{\'\i}nez-Delgado
          \inst{1}\fnmsep\thanks{Talentia Senior Fellow, dmartinez@iaa.es}
          \and
          Santi Roca-F\`abrega\inst{2,3,4}
          \and 
          Juan Mir\'o-Carretero\inst{3}
          \and 
          Maria~Angeles G\'omez-Flechoso\inst{3,4}
                    \and 
          Javier Román\inst{5,6,7}
                    \and 
          Giuseppe Donatiello  \inst{8}   
                     \and
          Judy Schmidt\inst{9}
          \and
          Dustin Lang\inst{10}
          \and
          Mohammad Akhlaghi \inst{11}
          }

   \institute{Instituto de Astrof\'isica de Andaluc\'ia, CSIC, Glorieta de la Astronom\'\i a, E-18080, Granada, Spain
   \and
Instituto de Astronomía, Universidad Nacional Autónoma de México, Apartado Postal 106, C. P. 22800, Ensenada, B. C., Mexico
\and
Departamento de F{\'\i}sica de la Tierra y Astrof{\'\i}sica, Universidad Complutense de Madrid, E-28040 Madrid, Spain
\and
Instituto de Física de Partículas y del Cosmos (IPARCOS), Fac. CC. Físicas, Universidad Complutense de Madrid, Plaza de las Ciencias, 1, E-28040 Madrid, Spain
\and
Kapteyn Astronomical Institute, University of Groningen, PO Box 800, 9700 AV Groningen, The Netherlands
\and
Instituto de Astrof\'isica de Canarias, c/ V\'ia L\'actea s/n, 38205 La Laguna, Tenerife, Spain 
\and
Departamento de Astrof\'isica, Universidad de La Laguna, E-38200 La Laguna, Spain 
\and
UAI - Unione Astrofili Italiani /P.I. Sezione Nazionale di Ricerca Profondo Cielo, 72024 Oria, Italy 
\and
Astrophysics Source Code Library, University of Maryland, 4254 Stadium Drive College Park, MD 20742, USA 
\and
Perimeter Institute for Theoretical Physics,
31 Caroline St N, Waterloo, Canada
\and
Centro de Estudios de Física del Cosmos de Aragón (CEFCA), Unidad Asociada al CSIC, Plaza San Juan 1, 44001 Teruel, Spain}
   \date{Received XXXX; accepted XXXX}

 
  \abstract
   {Tidal ring galaxies are observed rarely in the local universe due to their intrinsically transient nature. The tidal ring structures are the result of strong interactions between gas-rich stellar disks and smaller galactic systems and do not last longer than $\sim$500~Myr therefore, these are perfect scenarios where to find the debris of recently accreted dwarf galactic systems.}
   {Our goal is to study the low surface brightness stellar structures around the NGC~922 tidal ring galaxy, and to revise the hypothesis of its formation at the light of these new data.}
   {We present new deep images of the NGC~922 tidal ring galaxy and its surroundings from the DESI Legacy survey data and from our observations with an amateur telescope. These observations are compared with results from high-resolution N-body simulations designed to reproduce an alternative formation scenario for this peculiar galaxy.}
   {Our new observations unveil that the low surface brightness stellar tidal structures around NGC~922 are much more complex than reported in previous works. In particular, the formerly detected tidal spike-like structure at the northeast of the central galaxy disk is not connected with the dwarf companion galaxy PGC~3080368, which has been suggested as the intruder triggering the ring formation of NGC 922. The deep images reveal that this tidal structure is mainly composed by a fainter giant umbrella-like shape and thus it was formed from the tidal disruption of a different satellite. Using the broad-band $g$,  $r$ and $z$ DESI LS images,  we measured the photometric properties of this stellar stream, estimating a total absolute
   magnitude in $r$-band of M$_{r}$= -17.0  $\pm$ 0.03 magn and a total stellar mass for the stream between 6.9 --8.5 $\times$10$^{8}$ M$_{\odot}$. We perform a set of N-body simulations to reproduce the observed NGC922-intruder interaction, suggesting a new scenario for the formation of its tidal ring from the infall of a gas rich satellite around 150 Myr ago. Finally, our deep images also reveal a tidal shell around the dwarf galaxy PGC3080368, a possible fossil of a recent merger with a smaller satellite, which may suggest it is in its first infall towards NGC~922.}
   {}


   \maketitle
%

\section{Introduction}
\label{sec:intro} 

The widely accepted theory for galaxy formation and evolution proposes that galaxies grow through two main channels, the gas accretion through filaments, and mergers with other galactic systems \citep[e.g.,][]{Huillier2012}. These processes conduct to the formation of a zoo of galaxy morphologies that was described by the classical Hubble sequence \citep{Willett2013}. Furthermore, observations show that a non-negligible fraction of the Local Universe galaxies have peculiar morphologies not initially included in this classification \citep{Nair2010,Willett2013}. Among them, ring galaxies are particularly abundant \citep{Willett2013} as they are present in more than one fifth of spiral galaxies \citep{ButaCombes1996}. Their origin has been studied from the early development of numerical astronomy using N-body simulations \citep[e.g.,][]{Lynds&Toomre1976,Gerber1992,Hernquist&Weil1993}. Also experts on galactic dynamics used analytical approaches to unveil the origin of such structures \citep{StruckMarcell1990,KormendyKennicutt2004,RomeroGomez2007}. At the light of results from these many works the researchers agree that ring-like galactic systems can form through two mechanisms \citep{ButaCombes1996}, a secular process triggered by the presence of a galactic bar \citep{RomeroGomez2007}, and strong interactions with other galactic systems \citep{WuJiang2012}. Although both mechanisms produce similar morphologies, the properties of the resulting rings are well differentiated. Rings from secular origin are $\Theta$-shaped (O-Type), do not show expansion velocity, and are not or marginally star forming. Rings resulting from interactions (P-Type) show strong star formation within the ring and the central region of the galaxy, in most cases show a clear expansion velocity outwards, and present a variety of shapes.

\begin{figure*}
\centering
	\includegraphics[width=0.87\textwidth]{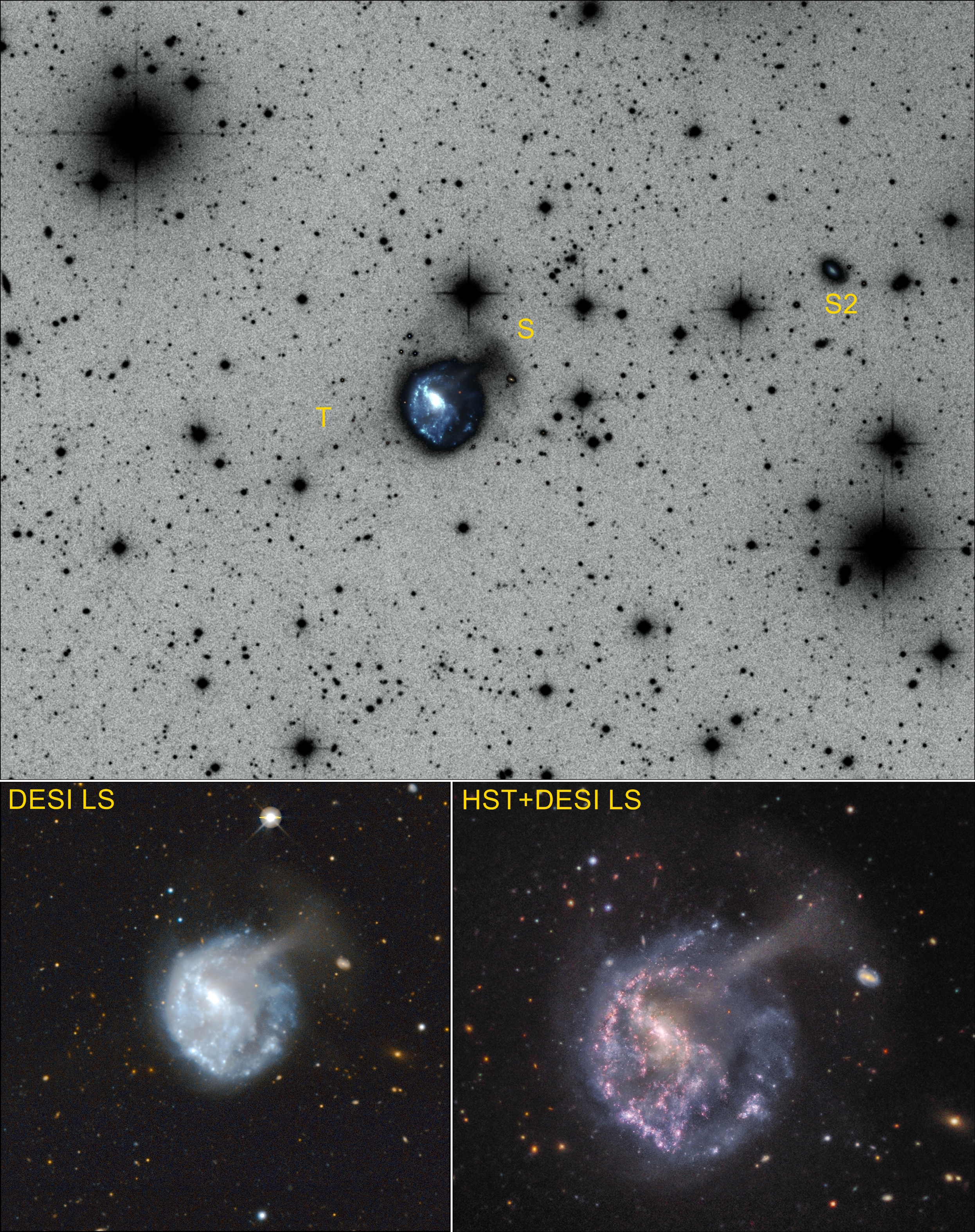}
    \caption{({\it Top}): The RC-60cm telescope luminance-filter image including NGC 922, its stellar tidal stream (which two components are marked with labels T and S) and the PG3080368 dwarf galaxy (marked with label S2). ({\it Bottom left}): DESI LS color image cutouf of NGC 922 and its stream obtained with {\it legacypipe} as described in Sec.~\ref{sec:data1} . ({\it Bottom right}): A zoomed, higher resolution view of the star formation regions of the NGC~922 disk and the stellar stream from a composition of public available {\it Hubble Space Telescope} images and the DESI Legacy survey data used in this work. }
    \label{fig:1}
\end{figure*}

Although collisions during minor mergers are common in the CDM paradigm, most collisional rings at the low-z universe are of secular origin.  In fact, the statistics of observed and simulated ring galaxies show that the presence of ring-like galaxies by collisions (Tidal Ring Galaxies, TRG hereafter) is specially relevant at high redshift when mergers dominate the galaxy growth \citep{ElagaliLagos2018}, while the secular processes are prevalent at lower-z and in low-density environments were galaxies tend to be less perturbed. This agrees well with the most recent observations of collisional rings in galaxies of the local universe \citep[e.g., CSRG and CZ2-CNRG][]{Buta1995,Buta2017} that according to \citet{Madore2009} show a lower limit of only the 0.001\%. In \citet{Madore2009} the authors also state that if assuming a constant minor mergers rate it is expected that in the local volume the majority of galaxies should have suffered a collisional ring formation over a Hubble time \citep[see also ][]{TheysSpiegel1977}. Therefore, the low number of detections is due to the short dynamical time of its formation, evolution, and decay.\\

After several years of research many studies showed that the TRGs are produced by an impulsive interaction between a small galaxy (intruder) in an almost radial orbit towards the gaseous disk of a larger galaxy \citep[e.g., ][]{Madore2009}. The short but strong interaction of the intruder with the disk of the larger galaxy make it to contract first and expand later resulting in an outwardly propagating wave \citep{Lynds&Toomre1976,Hernquist&Weil1993}. This compression wave triggers star formation in an expanding ring that can reach velocities up to $\sim$113~km~s$^{-1}$ just after the collision \citep[ $\sim$50~Myr, e.g., Arp 147][]{Fogarty2011}, slowing down to $\sim$50~km~s$^{-1}$ for older rings \citep[ $\sim$200~Myr, e.g., Cartwheel galaxy][]{Higdon1996}. Properties of the ring and the survival of the bar/bulge in the center of the disk are determined by the mass of the intruder and its impact parameter with respect to the center of the disk \citep{GerberLambBalsara1996,MadoreNelsonPetrillo2009,Fiacconi2012,ElagaliWong2018}. All models and simulations showed that these star forming rings only lasts for 0.2$-$0.5~Gyr \citep{Wong2006,Pellerin2010,Renaud2018,ElagaliLagos2018} and became only marginally observable up to 0.7~Gyr after the collision \citep{Wu2015}, which makes them transient features.  The identification of the intruder can also be challenging. Interacting dwarf galaxies can be quickly disrupted and mixed with the central regions of the TRG \citep{Madore2009}, either in their first or second pericenter  \citep{Wu2015}. In this regards \citet{ElmegreenElmegreen2006} showed that some observed TRGs have no obvious companions and that this can be due that observations are not deep enough or because they have already merged with the main galaxy. Some works also pointed out the possibility that multiple perturbers, or multiple interactions by a single intruder can generate prominent rings.

\begin{figure}
	\includegraphics[width=1.00\columnwidth]{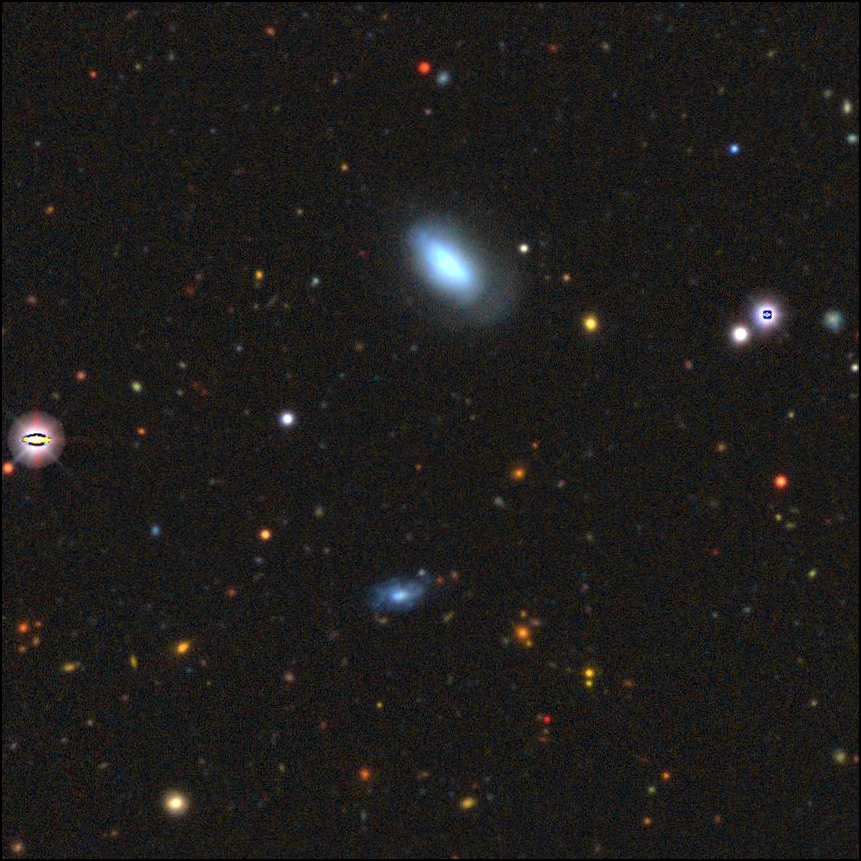}
    \caption{DESI Legacy image cutout of the compact dwarf galaxy PGC~3080368, the S2 intruder galaxy in \citet{Wong2006} intruder. This deep image reveals an outer shell in its Southwest side, also detected in the amateur data shown in Fig.~\ref{fig:1} ({\it top panel)}. The blue smaller galaxy to the South of S2 is possibly a background distant galaxy. The field-of-view is about 4 $\times$ 4 arcmin. North is up, East is left. }
    \label{fig:2}
\end{figure}

From an observational point of view the most prominent example of a TRG is the Cartwheel galaxy. This system has been studied and simulated by many research groups \citep[e.g.,][]{Charmandaris1999,Mayya2005,Barway2020}, which helped to settle down the picture of the formation and evolution of such systems. Another example of a Cartwheel-like system is the less studied NGC~922. \citet{Wong2006} proposed that the nearby compact dwarf PGC3080368 (named S2 in their work) could be the intruder that generated the observed TRG morphology on the NGC~922. In their work, the authors present an N-body simulation showing that an off-center impact of a point mass on the NGC~922-like disk galaxy can generate a TRG. However, their models can not fully explain the formation and morphology of the detected stellar plume around the NGC~922 \citep{Pellerin2010}. More recently, HI observations of the NGC~922 and its outskirts have found an HI tail that is not aligned with the S2 orbit neither with the stellar plume \citep{ElagaliWong2018}. This result suggest that other interaction than the one proposed by \citet{Wong2006} occurred to the NGC~922. In \citet{ElagaliWong2018} the authors have not found the HI bridge predicted by all hydrodynamical models of similar interacting systems that should connect the S2 and the NGC~922. In addition, although showing a Cartwheel-like morphology in observations in the optical, this system does not show the single drop-out TRG typical morphology not in HI neither in the stellar component as predicted by simulations  \citep{ElagaliLagos2018,Renaud2018}. All these inconsistencies found when comparing data with results from numerical experiments points towards a more complex formation scenario than the one proposed by \citet{Wong2006}.

In this paper, we present new deep images of the NGC~922 tidal ring galaxy and its surroundings from the DESI Legacy survey data and from our observations with an amateur telescope. These observations are compared with results from high-resolution N-body simulations designed to reproduce an alternative formation scenario for this peculiar galaxy. This paper is organized as follows. In Sec.~\ref{sec:observations} we describe the observations we use in this work to support our new hypothesis on the formation of the TRG NGC~922. In Sec.~\ref{sec:results} we present our main results obtained after analysing the observations. The N-body models used in this work are described in Sec.~\ref{sec:hydronbody}. Finally in Sec.~\ref{sec:4.3} and \ref{sec:conc} we discuss our results and present our conclusions.


\section{OBSERVATIONS AND DATA REDUCTION}\label{sec:observations} 

\subsection{Stan Watson Observatory South RC-60cm telescope}
\label{sec:data2} 
A wide-field deep image of NGC~922 and its tidal stream was obtained at the Stan Watson Observatory South (Dark Sky New Mexico, USA) with an 60 cm aperture $f/6.7$ Planewave CDK Ritchey–Chrétien telescope. We used a SBIG STX16803 CCD camera was that provided a pixel scale of 0.46$\arcsec$\,pixels$^{-1}$ over a $32.0\arcmin \times 32.0\arcmin$ field of view. We obtained a set of 40 individual 900 second  images with an Astrodon Gen2 Tru-Balance E-series luminance filter\footnote{This filter transmits from $400\lesssim \lambda \ (\mathrm{nm})\lesssim 700$, and broadly covers the $g$ and $r$ bands.}  over several nights between 2021 November 25--29 by remote observations. 
Each individual exposure was reduced following standard image processing procedures for dark subtraction, bias correction, and flat fielding \citep{2010AJ....140..962M}. The images were combined to create a final co-added luminance-filter image with a total exposure time of 36000\,s (see Fig.~\ref{fig:1} {\it top panel}).

\subsection{DESI Legacy surveys imaging data}
\label{sec:data1} 
NGC~922 is one of the target galaxies of the {\it Stellar Stream Legacy Survey} \citep{{MartinezDelgado2021arXiv}}, which is carrying on a systematic search for stellar tidal streams around massive galaxies of the local Universe. The imaging source for this project is the DESI Legacy Imaging Surveys (DESI LS), that compile optical data in three optical bands ($g$, $r,$ and $z$) obtained by three different imaging projects: The DECam Legacy Survey (DECaLS), the Beijing-Arizona Sky Survey (BASS), the Mayall $z$-band Legacy Survey (MzLS) \citep{2019ApJS..245....4Z, 2019AJ....157..168D} and re-reduced public DECam data from the DES \citep[][]{2018ApJS..239...18A}. 

In the bottom-left panel of Fig.~\ref{fig:1}we show an image cutout centred on NGC~922 obtained by coadding images of this galaxy taken by the DES \cite[]{2018ApJS..239...18A} using the DECam. These data were reprocessed using the \textsc{legacypipe} software of the DESI LS \citep[see e.g. Fig. 2 in ][]{MartinezDelgado2021arXiv}. In short, each image of this survey including NGC~922 was astrometrically calibrated to Gaia-DR2 and photometrically calibrated to the Pan-STARRS PS1 survey. Then they are subsequently resampled to a common pixel grid and summed with inverse-variance weighting.  

The depth of the DESI LS images in each band has been determined by calculating the surface brightness limit following the standard method of \cite{Roman2020}, i.e. the surface brightness corresponding to $3\sigma$ of the signal in the non detection areas of the image for a 100 arcsec$^2$ aperture \citep{MartinezDelgado2021arXiv}.  This yield 29.15, 28.73 and 27.52 mag\,arcsec$^{-2}$ for the $g$, the $r$ and the $z$ passbands, respectively. 

\subsection{Photometry of the NGC~922 stream}\label{sec:photometry} 

The photometry of the stellar stream around NGC 922 in the $g$, $r$, and $z$-bands has been derived with the {\it GNU Astronomy Utilities} (Gnuastro)\footnote{\url{http://www.gnu.org/software/gnuastro}} using resulting coadded image cutout of this galaxy obtained from the DESI LS data (see Sec.\ref{sec:data1}). The measurements have been carried out with Gnuastro's {\sc MakeCatalog} on the basis of the sky-subtracted image generated by Gnuastro's {\sc NoiseChisel} \citep{Akhlaghi2015,Akhlaghi2019}.  

In addition to the photometry of the NGC~922 stream obtained from this custom Legacy Surveys cutout image,  we also fitted a Sérsic model to the disk of NGC~922 in order to subtract spurious flux where the stream is located. Although this modeling does not contain structural information of NGC 922 given its irregular morphology, it is useful to reduce systematic errors and provide the cleanest possible photometry in external regions of the disk, where the stream stands out and is susceptible to analysis. Since the galaxy model used for the subtraction was not fully optimized for this type of irregular spiral galaxy,  
we only use it  in our photometric analysis to quantify the possible contamination of the outer disk of the host galaxy on the surface brightness, colour and luminosity measurements on the stream, as it is discussed in Sec.~\ref {sec:pholometry1}

\section{RESULTS}
\label{sec:results} 

\subsection{The umbrella-like stellar stream of NGC~922} \label{sec:progenitor}

\begin{figure}
  \includegraphics[width=0.95\columnwidth]{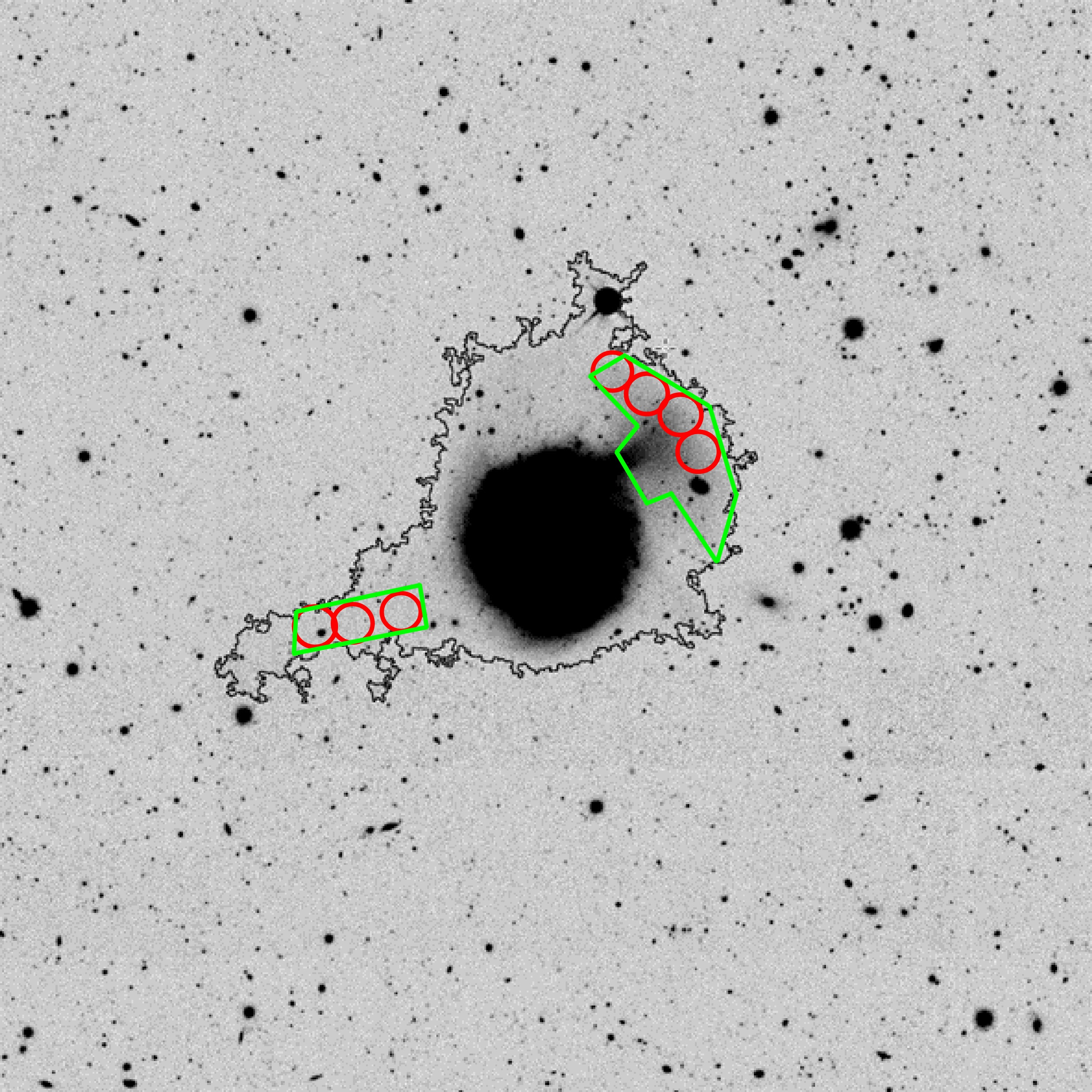}
  \caption{Photometry measurement method for the stellar stream around  NGC~922. The apertures where the photometry parameters, surface brightness and colours are measured both for the {\it shell} and the {\it tail} parts of the stream are highlighted in {\it red}. The polygonal apertures ({\it green}) indicate the parts of the image used to measure the magnitude of the stream. The contour of the detection, encompassing the galaxy and the stream, is highlighted by a {\it black} line. Here the input image has been warped to a pixel grid $8\times8$ courser than the original one, to highlight the low surface brightness signal.}  
  \label{fig-photometry}
\end{figure}

The deep images of NGC~922 detect  tidal structures in its outskirts at much lower surface brightness than previous studies and, thus, allow us to interpret their origin in more detail than in previous works. Fig.~\ref{fig:1} show that the halo of NGC~922 contains a giant, complex low surface brightness umbrella-like substructure connected to the aforementioned bright, collimated tidal plume detected in many previous studies (see Sec.~\ref{sec:intro}) and also visible in the HST data of this galaxy (Fig.~\ref{fig:1}, {\it bottom-right panel}).  Interestingly, both amateur (Fig.~\ref{fig:1}, {\it top panel}) and DESI LS (Fig.~\ref{fig-photometry}) images also reveal a very faint, tail-like feature in the southeast side of NGC 922 (marked with T in Fig.~\ref{fig:1} {\it top panel}), which seems to be the extension of the plume in the other side of the disk, following the path of the radial orbit of the disrupted satellite (see Sec.~\ref{sec:hydronbody}).  Thus,  our observations suggests that all these tidal  structures were formed by a different satellite than the nearby compact dwarf (named S2) proposed by \citet{Wong2006}, which appears clearly isolated and with no signature of a tidal tail emanating in the direction of NGC 922. Because of their shallower data, they assumed that the brighter tidal plume, roughly in the direction of the S2 dwarf,  was the only signature of a recent strong interaction of those galactic systems.

The overall tidal structure observed around NGC 922 is  quite similar  to that of the well-known  umbrella-like galactic systems \citep[e.g., NGC 4651,][]{Foster2014} are the consequence of the recent tidal disruption of a satellite galaxy orbiting in a radial orbit in the main galaxy outskirts.  Our simulations agree with these previous works (see Fig.~\ref{fig:simuNGC922}). Therefore, we state that there is no causal connection between the S2 and the formation of this low surface brightness structure.

Our deep amateur image also includes the S2 dwarf galaxy, clearly visible at the right upper side of our amateur image (marked with S2 in Fig.~\ref{fig:1} {\it top panel}). This kind of systems is expected to suffer a strong star formation burst after interacting with a central galaxy, followed by a fast depletion of gas as a consequence of the gas consumption in star formation and the clearance by supernovae feedback. Less than a Gyr after the interaction the satellite galaxy usually shows a low SFR and so an old stellar population \citep{Wetzel2013}. However, S2 looks very bright in the GALEX UV images, thus, it is actively forming stars. This result also disfavors the hypothesis that it interacted with NGC~922 to produce the observed tidal ring. In addition, the low surface brightness features detected in our image suggest that this dwarf galaxy holds its own tidal structure. Fig.~\ref{fig:2} shows an image cutout from the DESI LS data centered in this galaxy, showing the presence of a strong shell around S2.  This shell-like features look like very similar to those reported in deep images of the outskirts of the Large Magellanic Cloud \citep{Besla2016}, possibly originated by its interaction with the Small Magellanic Cloud. This S2 tidal features gives additional support that this satellite cannot be the intruder that generated the TRG, since a low-mass system like S2 should not hold such an unbound stellar structure after interacting with another galactic system.  The detection of this structure could also be a direct proof that dwarf satellites suffer similar accretion events as their larger counterparts. Although more data are needed this may also be a sign that the S2 it just in its first infall towards NGC~922.

\subsection{Stream surface brightness and colours}
\label{sec:pholometry1} 

The photometry measurements for the stream were carried out in circular apertures placed on clearly detected parts of the stream. Regions where the stream surface brightness could be contaminated by the outer disk of NGC~922 were avoided. Figure \ref{fig-photometry} illustrates the photometry measurement approach used in our analysis. After subtracting the sky from the input images, the first step is to perform the detection of the signal. Then all foreground and background objects, identified as {\it clumps} in Gnuastro's {\sc Segment} program \citep{Akhlaghi2015,Akhlaghi2019}, were masked and apertures were placed on the resulting image to measure the photometry parameters (e.g., surface brightness and colours). The {\it shell} on the NE side of the disk of NGC 922 (labelled as  S in Fig.\ref{fig:1}) and the {\it tail}  on the SW side (labelled as T in Fig.\ref{fig:1}) have been measured separately. For comparison, the photometry parameters have also been measured in an aperture placed on the galaxy. 

The results for the average surface brightness and colours along with the corresponding errors are given in Table~\ref{tab-photometry}, computed by Gnuastro's {\sc MakeCatalog}\footnote{\url{https://www.gnu.org/software//gnuastro/manual/html_node/MakeCatalog.html}}. The measurements have been averaged separately for the {\it shell}  and the {\it tail} described above. The surface brightness of the {\it tail} is approximately 2 mag\,arcsec$^{-2}$ fainter than in  the {\it shell}. The average colours differ somewhat between the {\it shell} and the {\it tail}, but are close enough (the shell colours are at 0.75 $\sigma$ for $(g-z)$ and 1.6 $\sigma$ for $(g-r)$ within the {\it tail} colour error distribution) to reinforce the notion that both parts belong to the same stream. The results are given for the DESI LS custom image, as the host galaxy-subtracted image suffers from over-subtraction and yields surface brightnesses up to 0.1  mag\,arcsec$^{-2}$ fainter than for the image without subtraction for the tail and up to 0.2  mag\,arcsec$^{-2}$ fainter for the shell, as expected, while the relative comparison between the {\it shell} and the {\it tail} remains the same. 

\begin{table*}
\centering                          
{\small
\caption{Surface brightnesses, colours and apparent magnitude  for  the tidal features detected around NGC 922. For the shell (S) and for the tail (T), the measurements listed are the average of the measurements on the circular apertures placed along the stream showed in  Fig.\ref{fig-photometry}. We also include surface brightness and color for  NGC 922 measured using a single aperture in its nominal center.}
\label{tab-photometry}
\addtolength{\tabcolsep}{-1.5pt}
\renewcommand{\arraystretch}{1.5}
\begin{tabular}{l c c c c c c c c}       
\hline\hline                 
 & <$\mu_{g}$> & <$\mu_{r}$> & <$\mu_{z}$> & <\textit{g}-\textit{r}$>_\textrm{stream}$ & <\textit{g}-\textit{z}$>_\textrm{stream}$ &
 m$_{g}$ & m$_{r}$ & m$_{z}$\\
       & [mag arcsec$^{-2}$] & [mag arcsec$^{-2}$] & [mag arcsec$^{-2}$] & [mag] & [mag] & [mag] & [mag] & [mag]\\
\hline                        
 Shell & 25.85 $\pm$ 0.02 & 25.34 $\pm$ 0.01 & 24.96 $\pm$ 0.03 & 0.53 $\pm$ 0.01 & 0.87 $\pm$ 0.005 & 16.80 $\pm$ 0.004 & 16.27 $\pm$ 0.004 & 15.93 $\pm$ 0.003  \\
 Tail  & 27.92 $\pm$ 0.08 & 27.27 $\pm$ 0.07 & 27.04 $\pm$ 0.16 & 0.61 $\pm$ 0.05 & 0.84 $\pm$ 0.04 & 19.95 $\pm$ 0.04 & 19.35 $\pm$ 0.03 & 19.11 $\pm$ 0.03 \\
NGC~922  & 22.59 $\pm$ 0.003 & 22.23 $\pm$ 0.003 & 22.21 $\pm$ 0.004 & 0.36 $\pm$ 0.003 & 0.58 $\pm$ 0.004 & & & \\ 
\hline                                   
\end{tabular}
}
\end{table*}

\subsection{Luminosity and Stellar Mass}
\label{sec:pholometry2} 

Unfortunately, it is not possible to derive the total luminosity of the NGC~922 stream,  mainly because of some parts of the stream are hidden (or overlapping) the host disk or are too faint to be detected due to the surface brightness limit of our images. However, it is worth attempting to approximate it in order to be able to properly constrain the stellar mass of its progenitor and to better understand the impact of its interaction with NGC~922. 
With this purpose, the apparent magnitude and total luminosity of the stream of the stream was measured using larger apertures covering as much area as possible of their different detected pieces. Two separate polygonal apertures were placed overlapping  the {\it shell} and the {\it tail} respectively, as depicted in Figure \ref{fig-photometry}. The measured apparent magnitudes are given in Table~\ref{tab-photometry}.

From the apparent magnitudes given in Table ~\ref{tab-photometry}
the NGC~922 distance \citep[d=43.1 Mpc from][]{Meurer2006}, and the Galactic extinction obtained from {\it NASA/IPAC Extragalactic Database}\footnote{\url{https://ned.ipac.caltech.edu/extinction_calculator}}), we derive an absolute magnitude of M$_{g}$= -16.5 $\pm$ 0.04 magn,  M$_{r}$= -17.0  $\pm$ 0.03 magn and M$_{z}$= -17.3 $\pm$ 0.03 magn for the whole stream (shell plus tail). We compute the luminosity of these two parts of the stream by using the solar absolute magnitudes for the $g$, the $r$ and the $z$ passbands from \citet{willmer2018}. We obtain the values of  4.14$\times$10$^8$ L$_{\odot}$ for the g-passband, 4.42$\times$10$^8$ L$_{\odot}$ for the r-passband and 5.35$\times$10$^8$ L$_{\odot}$ for the z-passband. We calculate the mass--to--light ratio (M/L$_{\lambda}$) from the 3 colours measured for the stream (see Section \ref{sec:pholometry1}), using the correlations between {\it SDSS ugriz} colors and {\it SDSS/2MASS} M/L ratios given in  \citet{2003ApJS..149..289B} (i.e., the coefficients given in Table {\it A7} of their paper). Following this method,  we obtain an estimate for the stellar mass of the stream progenitor between $6.87\times 10^8$  M$_{\odot}$ and $8.51\times 10^8$ M$_{\odot}$. The ratio between the stellar masses of the dwarf progenitor and the host spiral galaxy is key to determine the kind of tidal interaction between them and to characterize the resulting merger event. Considering the stellar mass of NGC922 is 5.47$\times$9$^8$ M$_{\odot}$ \citep {Wong2006}, this yield a (stellar) mass merger ratio between ${0.13}$ and ${0.16}$.

\section{N-BODY SIMULATIONS}\label{sec:hydronbody} 

\begin{figure}
\centering
	\includegraphics[width=0.467\textwidth]{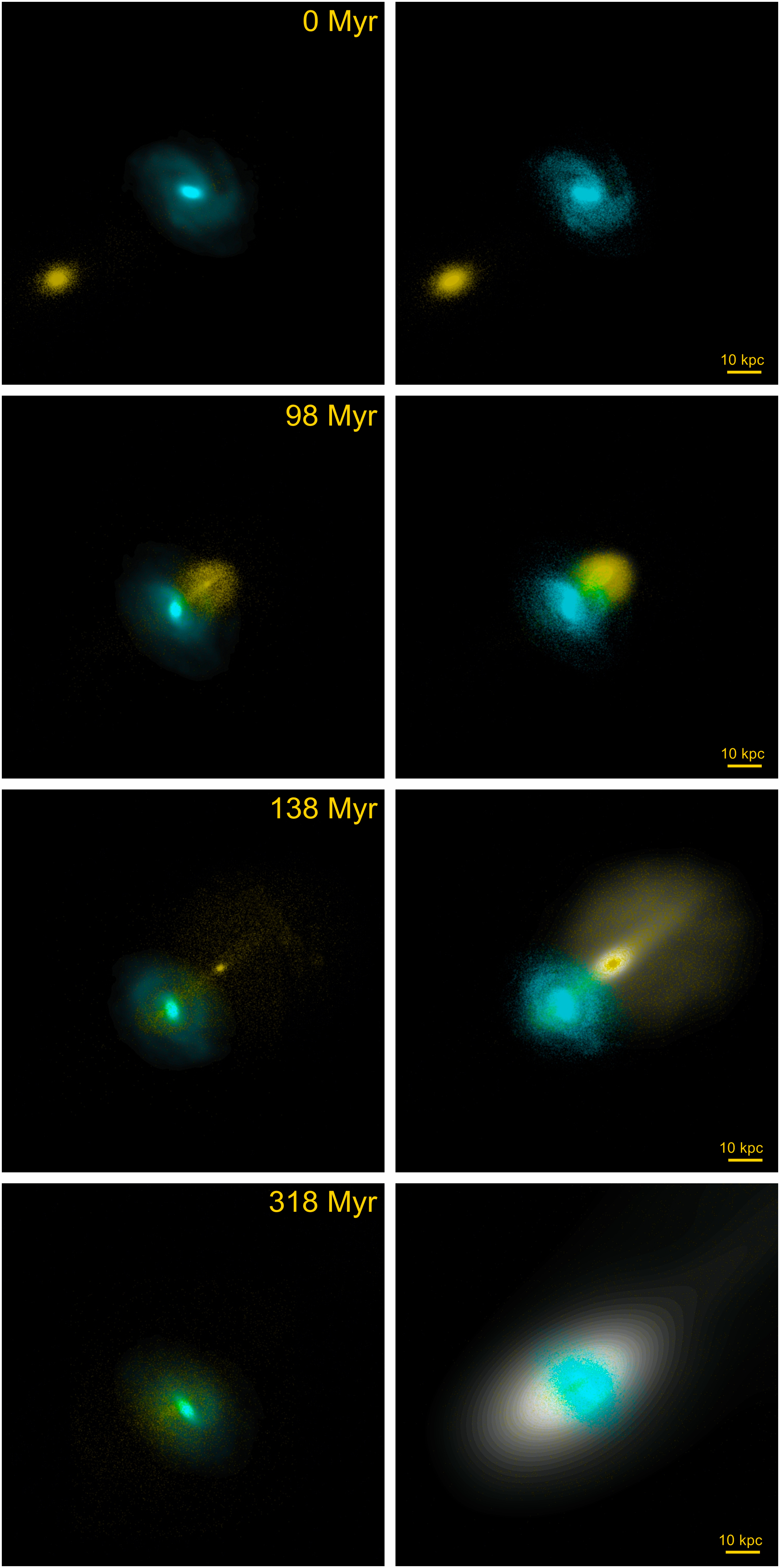}
    \caption{Four snapshots of the N-body simulation of a NGC922-like Tidal Ring Galaxy formation. Left panels: Focusing on the tidal ring formation within the main galactic system (cyan). Right panel: Putting the focus on the intruder evolution (yellow) and on the formation of tidal shells. The zero point of the time progression labeled in the left panels is arbitrary and do not correspond to the initial conditions of the simulation.}
    \label{fig:simuNGC922}
\end{figure}

We performed a set of N-body simulations using the ART code \citep{Kravtsov1997,Colin2010} starting with initial conditions designed to reproduce the NGC~922-intruder interaction that generated the observed stellar plume \citep{Pellerin2010}. We used a single dark matter particle species with a mass of 2.5$\times$10$^4$~M$_{\odot}$, which is the same we set as the maximum star particle mass. The spatial resolution (one AMR cell side) is 40~pc.\\
The simulated system includes a NGC~922-like and a compact dwarf galaxies, both inside a box of 1~Mpc/h side. The NGC~922-like galaxy is simulated as a stellar exponential disk embedded in a NFW dark matter halo. In Tab.~\ref{tab:Nbody} we show the parameters we used to generate the stellar disk and the dark matter halo profiles. The initial conditions of the collisionless components have been obtained using the Jeans equation moments method as introduced by \citet{Hernquist1993}. We chose these parameters to be consistent with results from our own observations of the NGC~922 system and also the ones by \citet{Wong2006,Pellerin2010,ElagaliWong2018}. 
In contrast with previous works, we simulated the compact dwarf system as an extended distribution of particles, which allowed us to reproduce the observed stellar stream. The initial condition of this system is a simple stellar structure that follows a compact NFW profile (see Tab.~\ref{tab:Nbody}).\\
We located the compact dwarf at (1.0,0.0,100.0)~kpc, being the NGC~922-like system at (0.0,0.0,0.0)~kpc, and with a relative velocity of (0.0,0.0,-500)~km/s. The initial location and velocity of the intruder have been chosen to produce an off-center collision, this following the works by \citet{Wong2006} and \citet{Renaud2018}.
We evolved the system for 2~Gyr which allows us to study the intruder's orbit from its first infall to an almost complete disruption (see Fig.~\ref{fig:simuNGC922}).

\begin{table*}
\centering                          
\caption{Initial setup of the NGC~922-like and compact dwarf galactic systems.}. 
\label{tab:Nbody}     
\begin{tabular}{c c c r}       
\hline\hline
 Properties & NGC922-like & Compact Dwarf & \\
\hline                        
M$_{tot}$ [M$_{\odot}$]  & 1.22$\times$10$^{11}$ & 5$\times$10$^{9}$ & Total mass\\
M$_*$ [M$_{\odot}$]  & 2.5$\times$10$^{10}$ & 8$\times$10$^{8}$ & Stellar mass\\
R$_d$ [kpc] & 3.14 & & Stellar disk exponential scale length\\
z$_d$ [kpc] & 0.20 & & Stellar disk scale height\\
R$_{dtrunc}$ [kpc] & 14.30 & & Stellar disk truncation radius\\
Q  & 1.2 & & Toomre parameter\\
C$_{NFW}$ & 15 & 10 & Concentration parameter\\
R$_s$ [kpc] & 8.27 &  2.4 & Halo scale radius\\
R$_{htrunc}$ [R$_s$] & 25.0 & 2 & Halo truncation radius in R$_s$ units\\
\hline                                   
\end{tabular}
\end{table*}

\section{DISCUSSION}
\label{sec:4.3}
In this section we analyse two scenarios for the formation of the NGC~922's tidal ring structure. First, we study the currently accepted scenario where the S2 dwarf galaxy is the intruder. Later, we propose a new scenario where the intruder is a much closer satellite that is now almost completely disrupted, that also generated the umbrella-like stellar stream, and the detected HI tail.\\

In the scenario proposed by \citet{Wong2006}, the S2 strongly interacted with NGC~922 and quickly moved to its current location. If we assume NGC~922 and S2 are roughly at the same distance from the Sun \citep[i.e., 43.1~Mpc according to][]{Meurer2006}, this means that S2 moved at a minimum distance of $\sim$100~Kpc from the NGC~922's center.
If S2 is not gravitationally bound to the central galaxy (parabolic orbit), the time interval between two positions $d_1$ and $d_2$ of the trajectory is roughly $\Delta t\sim (d_2^{3/2}-d_1^{3/2})/\sqrt{4.5 G M} $. Therefore, assuming a close encounter between NGC~922 and S2 and no mass-loss, dynamical friction or other slow-down processes, the S2 dwarf would need, at least, 0.7 Gyr to reach 100 Kpc distance from the NGC~922 center. If S2 is gravitationally bound, this time interval would be even larger.
All models and simulations showed that the star forming rings associated with the tidal ring structures only lasts for 0.2$-$0.5~Gyr \citep{Wong2006,Pellerin2010,Renaud2018,ElagaliLagos2018} and became only marginally observable up to 0.7~Gyr after the collision \citep{Wu2015}. The NGC~922's star forming ring is clearly visible (see Fig.~\ref{fig:1}), so it is in an early stage of its evolution. Both results are not compatible, therefore we state that this first scenario should be put in doubt. In addition, as discussed in Sec.~\ref{sec:progenitor}, the S2 dwarf holds its own well defined tidal shell structure, a feature that should not exist if the dwarf had a strong close encounter with NGC~922.
We also warn the reader that although \citet{Wong2006} reproduces such scenario using pure N-body simulations, in their models they simulate the intruder as a point-mass, thus, they do not capture most of the tidal effects that affect its evolution. In this simplified model the effects over the central galaxy structure are also artificially magnified. 
In our models, the intruder has internal structure and it still induces a tidal ring on the central galaxy. The largest difference between our model and the model of \citet{Wong2006} is the fate of the intruder. In agreement with previous models \citep[see e.g.,][]{Foster2014} the intruder is partially or totally disrupted and generates a variety of tidal structures around the central galaxy (see Fig.~\ref{fig:simuNGC922}). We also want to notice that the TRG morphology obtained in our simulations lasts for less than 500~Myr after the intruder's first pericenter, as predicted by previous works. This result reinforces the hypothesis that a recent interaction is responsible for the tidal ring structure.
Finally, it is also important to mention that in most of the studied TRG the authors detect HI tails in the direction of the intruder \citep[see e.g.,][]{Higdon1996}. In \citet{ElagaliWong2018} the authors found an HI tail but it points in other direction extending towards north of the disk instead of east, and for $\sim$0.7~arcmin (8~Kpc). This result is also inconsistent with the simple collisional scenario proposed by \citet{Wong2006} for the NGC~922 formation. \\

The discovery of the new stream morphology (see Sec~\ref{sec:progenitor}) and the results from our N-body models (see Fig.~\ref{fig:simuNGC922}) brought us to a new interpretation for the origin of the NGC~922 morphology. From an observational point of view we detect that a satellite merged recently with the host producing two shells one evident in the north-west, another more diffuse in the south-east, with a total mass of $\sim$6.9-8.5$\times$10$^{8}$~M$_{\odot}$ \citep[this work, in agreement with resuts by][]{Pellerin2010}. We expect that stars of this dwarf satellite are now partially mixed with stars of the central. If the mixing truly occurred this would be reflected by a bi-modal distribution in metallicity. In \citet{Kouroumpatzakis2021} the authors found evidences of two metallicity populations in NGC~922, result that is consistent with the new scenario presented here. Additionally, as the NGC~922 is a relatively low-mass galaxy, the infalling satellites do not suffer strong ram-pressure stripping and bring fresh HI gas that fuels star formation in the main system. The recent infall of a gas rich satellite also explains why NGC~922 has M$_{HI}$/M$_*\sim$0.32 that is much higher than the one observed in galaxies with similar masses ($\sim$0.13)\citep[see][]{ElagaliWong2018}. The presence of these tidal shells can also explain the strange kinematic signatures detected in the external regions of the host and that made some authors propose that the system was already perturbed before the formation of the tidal ring \citep[e.g.,][]{Renaud2018}.
Our N-body models also support this scenario that contradicts the formerly accepted one, where S2 was the intruder (see Fig.~\ref{fig:simuNGC922}).

\section{CONCLUSIONS}\label{sec:conc}

We present a new merger history for the formation of the tidal ring galaxy NGC~922 after obtaining a set of deep images revealing its low-surface brightness outskirts. The main conclusions of this work are the following:
\begin{itemize}
\item We report the presence of an umbrella-like tidal structure where previous observations proposed the presence of a connecting tidal tail with a nearby dwarf galaxy. 
\item The absence of such a tidal tail connecting NGC~922 with the dwarf companion S2 (PGC~3080368), both in stars or in HI, may discard the hypothesis of the latter being the trigger of the tidal ring formation. Recent works also raised the question of whether a scenario as simple as a single off-shot could explain the many peculiar properties of this system.
\item We propose a new scenario for the formation of the tidal ring in NGC~922, that is a merger with an already totally or partially stripped dwarf galaxy. This new scenario explains many of the previous peculiar results like high velocities in the outskirts of the galaxy, the presence of two metallicity populations, and the relatively recent formation of the ring, if the expansion velocities and lifetimes predicted by numerical models are correct. Also a recent gas rich wet merger can explain the observed high HI mass fraction of the galactic system.
\item Finally we present the discovery of a tidal structure around the dwarf galaxy S2, which holds its own tidal shell structure. This, together with its high star formation rate shown by the GALEX-UV data, could be an evidence that the system is in its first in-fall towards NGC~922. This result can be confirmed by future estimations of the radial velocity and proper motions of this satellite.
\end{itemize}
Our work also showed as deep imaging of the TRG's properties, reaching a enough surface brightness regime for detecting the presence of giant tidal shells in their outskirts, can provide valuable information on the recent evolution of galaxies undergoing a minor merger. In the {\it Stellar Tidal Stream Survey} (Mart{\'\i}nez-Delgado et al. 2022, in preparation), we have detected several nearby spiral galaxies with both tidal ring and tidal shell features in different evolutionary stages. Our observations include  galaxies with recently formed tidal rings and single tenuous shells, others with well-developed tidal rings and a single shell, and also some with evidence of multiple tidal rings and shells. However, to make a systematic study of the relation between the formation of tidal rings and the almost radial minor mergers that generate tidal shells, it is fundamental to carry out a systematic survey of TRGs with a surface brightness limit fainter than 28 mag arcsec$^{-2}$. Once completed, this deep imaging survey combined with future space missions like EUCLID will provide us with enough data to fully characterize the TRG in the local volume and, thus, to better understand the processes that shape galaxies during and after a radial minor merger. This information will also open the door to study whether the Milky Way suffered such a process in its recent history.


\begin{acknowledgements}
 DMD acknowledges financial support from the Talentia Senior Program (through the incentive ASE-136) from Secretar\'\i a General de  Universidades, Investigaci\'{o}n y Tecnolog\'\i a, de la Junta de Andaluc\'\i a. DMD acknowledge funding from the State Agency for Research of the Spanish MCIU through the ``Center of Excellence Severo Ochoa" award to the Instituto de Astrof{\'i}sica de Andaluc{\'i}a (SEV-2017-0709) and project (PDI2020-114581GB-C21/ AEI / 10.13039/501100011033). SRF acknowledge financial support from the Spanish Ministry of Economy and Competitiveness (MINECO) under grant number AYA2016-75808-R, AYA2017-90589-REDT and S2018/NMT-429, and from the CAM-UCM under grant number PR65/19-22462. SRF acknowledges support from a Spanish postdoctoral fellowship, under grant number 2017-T2/TIC-5592. MAGF acknowledges financial support from the Spanish Ministry of Science and Innovation through the project PID2020-114581GB-C22. JR acknowledges support from the State Research Agency (AEI-MCINN) of the Spanish Ministry of Science and Innovation under the grant "The structure and evolution of galaxies and their central regions" with reference PID2019-105602GB-I00/10.13039/501100011033. JR also acknowledges funding from University of La Laguna through the Margarita Salas Program from the Spanish Ministry of Universities ref. UNI/551/2021-May 26, and under the EU Next Generation.
 M.A acknowledges the financial support from the Spanish Ministry of Science and Innovation and the European Union - NextGenerationEU through the Recovery and Resilience Facility project ICTS-MRR-2021-03-CEFCA.

This project uses data from observations at Cerro Tololo Inter-American Observatory, National Optical Astronomy Observatory, which is operated by the Association of Universities for Research in Astronomy (AURA) under a cooperative agreement with the National Science Foundation. We acknowledge support from the Spanish Ministry for Science, Innovation and Universities and FEDER funds through grant AYA2016-81065-C2-2. We also used data obtained with the Dark Energy Camera (DECam), which was constructed by the Dark Energy Survey (DES) collaboration. Funding for the DES Projects has been provided by
the U.S. Department of Energy, 
the U.S. National Science Foundation, 
the Ministry of Science and Education of Spain, 
the Science and Technology Facilities Council of the United Kingdom, 
the Higher Education Funding Council for England, 
the National Center for Supercomputing Applications at the University of Illinois at Urbana-Champaign, 
the Kavli Institute of Cosmological Physics at the University of Chicago, 
the Center for Cosmology and Astro-Particle Physics at the Ohio State University, 
the Mitchell Institute for Fundamental Physics and Astronomy at Texas A\&M University, 
Financiadora de Estudos e Projetos, Funda{\c c}{\~a}o Carlos Chagas Filho de Amparo {\`a} Pesquisa do Estado do Rio de Janeiro, 
Conselho Nacional de Desenvolvimento Cient{\'i}fico e Tecnol{\'o}gico and the Minist{\'e}rio da Ci{\^e}ncia, Tecnologia e Inovac{\~a}o, 
the Deutsche Forschungsgemeinschaft, 
and the Collaborating Institutions in the Dark Energy Survey. 
The Collaborating Institutions are 
Argonne National Laboratory, 
the University of California at Santa Cruz, 
the University of Cambridge, 
Centro de Investigaciones En{\'e}rgeticas, Medioambientales y Tecnol{\'o}gicas-Madrid, 
the University of Chicago, 
University College London, 
the DES-Brazil Consortium, 
the University of Edinburgh, 
the Eidgen{\"o}ssische Technische Hoch\-schule (ETH) Z{\"u}rich, 
Fermi National Accelerator Laboratory, 
the University of Illinois at Urbana-Champaign, 
the Institut de Ci{\`e}ncies de l'Espai (IEEC/CSIC), 
the Institut de F{\'i}sica d'Altes Energies, 
Lawrence Berkeley National Laboratory, 
the Ludwig-Maximilians Universit{\"a}t M{\"u}nchen and the associated Excellence Cluster Universe, 
the University of Michigan, 
{the} National Optical Astronomy Observatory, 
the University of Nottingham, 
the Ohio State University, 
the University of Pennsylvania, 
the University of Portsmouth, 
SLAC National Accelerator Laboratory, 
Stanford University, 
the University of Sussex, 
and Texas A\&M University.
Support for this work was provided by NASA through the NASA Hubble Fellowship grant \#HST-HF2-51466.001-A awarded by the Space Telescope Science Institute, which is operated by the Association of Universities for Research in Astronomy, Incorporated, under NASA contract NAS5-26555.
This work was partly done using GNU Astronomy Utilities (Gnuastro, ascl.net/1801.009) version 0.17. Work on Gnuastro has been funded by the Japanese MEXT scholarship and its Grant-in-Aid for Scientific Research (21244012, 24253003), the European Research Council (ERC) advanced grant 339659-MUSICOS, and from the Spanish Ministry of Economy and Competitiveness (MINECO) under grant number AYA2016-76219-P.
M.A acknowledges the financial support from the Spanish Ministry of Science and Innovation and the European Union - NextGenerationEU through the Recovery and Resilience Facility project ICTS-MRR-2021-03-CEFCA

\end{acknowledgements}

%
\bibliographystyle{aa} 
\bibliography{ref} 

\begin{thebibliography}{48}
\expandafter\ifx\csname natexlab\endcsname\relax\def\natexlab#1{#1}\fi

\bibitem[{{Abbott} {et~al.}(2018){Abbott}, {Abdalla}, {Allam}, \& {et
  al.}}]{2018ApJS..239...18A}
{Abbott}, T.~M.~C., {Abdalla}, F.~B., {Allam}, \& {et al.} 2018, \apjs, 239, 18

\bibitem[{{Akhlaghi}(2019)}]{Akhlaghi2019}
{Akhlaghi}, M. 2019, arXiv e-prints, arXiv:1909.11230

\bibitem[{{Akhlaghi} \& {Ichikawa}(2015)}]{Akhlaghi2015}
{Akhlaghi}, M. \& {Ichikawa}, T. 2015, \apjs, 220, 1

\bibitem[{{Barway} {et~al.}(2020){Barway}, {Mayya}, \&
  {Robleto-Or{\'u}s}}]{Barway2020}
{Barway}, S., {Mayya}, Y.~D., \& {Robleto-Or{\'u}s}, A. 2020, \mnras, 497, 44

\bibitem[{{Bell} {et~al.}(2003){Bell}, {McIntosh}, {Katz}, \&
  {Weinberg}}]{2003ApJS..149..289B}
{Bell}, E.~F., {McIntosh}, D.~H., {Katz}, N., \& {Weinberg}, M.~D. 2003, \apjs,
  149, 289

\bibitem[{{Besla} {et~al.}(2016){Besla}, {Mart{\'\i}nez-Delgado}, {van der
  Marel}, {Beletsky}, {Seibert}, {Schlafly}, {Grebel}, \& {Neyer}}]{Besla2016}
{Besla}, G., {Mart{\'\i}nez-Delgado}, D., {van der Marel}, R.~P., {et~al.}
  2016, \apj, 825, 20

\bibitem[{{Buta}(1995)}]{Buta1995}
{Buta}, R. 1995, \apjs, 96, 39

\bibitem[{{Buta} \& {Combes}(1996)}]{ButaCombes1996}
{Buta}, R. \& {Combes}, F. 1996, \fcp, 17, 95

\bibitem[{{Buta}(2017)}]{Buta2017}
{Buta}, R.~J. 2017, \mnras, 471, 4027

\bibitem[{{Charmandaris} {et~al.}(1999){Charmandaris}, {Laurent}, {Mirabel},
  {Gallais}, {Sauvage}, {Vigroux}, {Cesarsky}, \&
  {Appleton}}]{Charmandaris1999}
{Charmandaris}, V., {Laurent}, O., {Mirabel}, I.~F., {et~al.} 1999, \aap, 341,
  69

\bibitem[{{Col{\'\i}n} {et~al.}(2010){Col{\'\i}n}, {Avila-Reese},
  {V{\'a}zquez-Semadeni}, {Valenzuela}, \& {Ceverino}}]{Colin2010}
{Col{\'\i}n}, P., {Avila-Reese}, V., {V{\'a}zquez-Semadeni}, E., {Valenzuela},
  O., \& {Ceverino}, D. 2010, \apj, 713, 535

\bibitem[{{Dey} {et~al.}(2019){Dey}, {Schlegel}, {Lang}, \& {et
  al.}}]{2019AJ....157..168D}
{Dey}, A., {Schlegel}, D.~J., {Lang}, D., \& {et al.} 2019, \aj, 157, 168

\bibitem[{{Elagali} {et~al.}(2018{\natexlab{a}}){Elagali}, {Lagos}, {Wong},
  {Staveley-Smith}, {Trayford}, {Schaller}, {Yuan}, \&
  {Abadi}}]{ElagaliLagos2018}
{Elagali}, A., {Lagos}, C. D.~P., {Wong}, O.~I., {et~al.} 2018{\natexlab{a}},
  \mnras, 481, 2951

\bibitem[{{Elagali} {et~al.}(2018{\natexlab{b}}){Elagali}, {Wong}, {Oh},
  {Staveley-Smith}, {Koribalski}, {Bekki}, \& {Zwaan}}]{ElagaliWong2018}
{Elagali}, A., {Wong}, O.~I., {Oh}, S.-H., {et~al.} 2018{\natexlab{b}}, \mnras,
  476, 5681

\bibitem[{{Elmegreen} \& {Elmegreen}(2006)}]{ElmegreenElmegreen2006}
{Elmegreen}, D.~M. \& {Elmegreen}, B.~G. 2006, \apj, 651, 676

\bibitem[{{Fiacconi} {et~al.}(2012){Fiacconi}, {Mapelli}, {Ripamonti}, \&
  {Colpi}}]{Fiacconi2012}
{Fiacconi}, D., {Mapelli}, M., {Ripamonti}, E., \& {Colpi}, M. 2012, \mnras,
  425, 2255

\bibitem[{{Fogarty} {et~al.}(2011){Fogarty}, {Thatte}, {Tecza}, {Clarke},
  {Goodsall}, {Houghton}, {Salter}, {Davies}, \& {Kassin}}]{Fogarty2011}
{Fogarty}, L., {Thatte}, N., {Tecza}, M., {et~al.} 2011, \mnras, 417, 835

\bibitem[{{Foster} {et~al.}(2014){Foster}, {Lux}, {Romanowsky},
  {Mart{\'\i}nez-Delgado}, {Zibetti}, {Arnold}, {Brodie}, {Ciardullo},
  {GaBany}, {Merrifield}, {Singh}, \& {Strader}}]{Foster2014}
{Foster}, C., {Lux}, H., {Romanowsky}, A.~J., {et~al.} 2014, \mnras, 442, 3544

\bibitem[{{Gerber} {et~al.}(1992){Gerber}, {Lamb}, \& {Balsara}}]{Gerber1992}
{Gerber}, R.~A., {Lamb}, S.~A., \& {Balsara}, D.~S. 1992, \apjl, 399, L51

\bibitem[{{Gerber} {et~al.}(1996){Gerber}, {Lamb}, \&
  {Balsara}}]{GerberLambBalsara1996}
{Gerber}, R.~A., {Lamb}, S.~A., \& {Balsara}, D.~S. 1996, \mnras, 278, 345

\bibitem[{{Hernquist}(1993)}]{Hernquist1993}
{Hernquist}, L. 1993, \apjs, 86, 389

\bibitem[{{Hernquist} \& {Weil}(1993)}]{Hernquist&Weil1993}
{Hernquist}, L. \& {Weil}, M.~L. 1993, \mnras, 261, 804

\bibitem[{{Higdon}(1996)}]{Higdon1996}
{Higdon}, J.~L. 1996, \apj, 467, 241

\bibitem[{{Kormendy} \& {Kennicutt}(2004)}]{KormendyKennicutt2004}
{Kormendy}, J. \& {Kennicutt}, Robert~C., J. 2004, \araa, 42, 603

\bibitem[{{Kouroumpatzakis} {et~al.}(2021){Kouroumpatzakis}, {Zezas}, {Wolter},
  {Fruscione}, {Anastasopoulou}, \& {Prestwich}}]{Kouroumpatzakis2021}
{Kouroumpatzakis}, K., {Zezas}, A., {Wolter}, A., {et~al.} 2021, \mnras, 500,
  962

\bibitem[{{Kravtsov} {et~al.}(1997){Kravtsov}, {Klypin}, \&
  {Khokhlov}}]{Kravtsov1997}
{Kravtsov}, A.~V., {Klypin}, A.~A., \& {Khokhlov}, A.~M. 1997, \apjs, 111, 73

\bibitem[{{L'Huillier} {et~al.}(2012){L'Huillier}, {Combes}, \&
  {Semelin}}]{Huillier2012}
{L'Huillier}, B., {Combes}, F., \& {Semelin}, B. 2012, \aap, 544, A68

\bibitem[{{Lynds} \& {Toomre}(1976)}]{Lynds&Toomre1976}
{Lynds}, R. \& {Toomre}, A. 1976, \apj, 209, 382

\bibitem[{{Madore} {et~al.}(2009{\natexlab{a}}){Madore}, {Nelson}, \&
  {Petrillo}}]{Madore2009}
{Madore}, B.~F., {Nelson}, E., \& {Petrillo}, K. 2009{\natexlab{a}}, \apjs,
  181, 572

\bibitem[{{Madore} {et~al.}(2009{\natexlab{b}}){Madore}, {Nelson}, \&
  {Petrillo}}]{MadoreNelsonPetrillo2009}
{Madore}, B.~F., {Nelson}, E., \& {Petrillo}, K. 2009{\natexlab{b}}, \apjs,
  181, 572

\bibitem[{{Martinez-Delgado} {et~al.}(2021){Martinez-Delgado}, {Cooper},
  {Roman}, {Pillepich}, {Erkal}, {Pearson}, {Moustakas}, {Laporte}, {Laine},
  {Akhlaghi}, {Lang}, {Makarov}, {Borlaff}, {Donatiello}, {Pearson},
  {Miro-Carretero}, {Cuillandre}, {Dominguez}, {Roca-Fabrega}, {Frenk},
  {Schmidt}, {Gomez-Flechoso}, {Guzman}, {Libeskind}, {Dey}, {Weaver},
  {Schlegel}, {Myers}, \& {Valdes}}]{MartinezDelgado2021arXiv}
{Martinez-Delgado}, D., {Cooper}, A.~P., {Roman}, J., {et~al.} 2021, arXiv
  e-prints, arXiv:2104.06071

\bibitem[{{Mart{\'\i}nez-Delgado} {et~al.}(2010){Mart{\'\i}nez-Delgado},
  {Gabany}, {Crawford}, {Zibetti}, {Majewski}, {Rix}, {Fliri},
  {Carballo-Bello}, {Bardalez-Gagliuffi}, {Pe{\~n}arrubia}, {Chonis}, {Madore},
  {Trujillo}, {Schirmer}, \& {McDavid}}]{2010AJ....140..962M}
{Mart{\'\i}nez-Delgado}, D., {Gabany}, R.~J., {Crawford}, K., {et~al.} 2010,
  \aj, 140, 962

\bibitem[{{Mayya} {et~al.}(2005){Mayya}, {Bizyaev}, {Romano}, {Garcia-Barreto},
  \& {Vorobyov}}]{Mayya2005}
{Mayya}, Y.~D., {Bizyaev}, D., {Romano}, R., {Garcia-Barreto}, J.~A., \&
  {Vorobyov}, E.~I. 2005, \apjl, 620, L35

\bibitem[{{Meurer} {et~al.}(2006){Meurer}, {Hanish}, {Ferguson}, {Knezek},
  {Kilborn}, {Putman}, {Smith}, {Koribalski}, {Meyer}, {Oey}, {Ryan-Weber},
  {Zwaan}, {Heckman}, {Kennicutt}, {Lee}, {Webster}, {Bland-Hawthorn},
  {Dopita}, {Freeman}, {Doyle}, {Drinkwater}, {Staveley-Smith}, \&
  {Werk}}]{Meurer2006}
{Meurer}, G.~R., {Hanish}, D.~J., {Ferguson}, H.~C., {et~al.} 2006, \apjs, 165,
  307

\bibitem[{{Nair} \& {Abraham}(2010)}]{Nair2010}
{Nair}, P.~B. \& {Abraham}, R.~G. 2010, \apjs, 186, 427

\bibitem[{{Pellerin} {et~al.}(2010){Pellerin}, {Meurer}, {Bekki}, {Elmegreen},
  {Wong}, \& {Knezek}}]{Pellerin2010}
{Pellerin}, A., {Meurer}, G.~R., {Bekki}, K., {et~al.} 2010, \aj, 139, 1369

\bibitem[{{Renaud} {et~al.}(2018){Renaud}, {Athanassoula}, {Amram}, {Bosma},
  {Bournaud}, {Duc}, {Epinat}, {Fensch}, {Kraljic}, {Perret}, \&
  {Struck}}]{Renaud2018}
{Renaud}, F., {Athanassoula}, E., {Amram}, P., {et~al.} 2018, \mnras, 473, 585

\bibitem[{{Rom{\'a}n} {et~al.}(2020){Rom{\'a}n}, {Trujillo}, \&
  {Montes}}]{Roman2020}
{Rom{\'a}n}, J., {Trujillo}, I., \& {Montes}, M. 2020, \aap, 644, A42

\bibitem[{{Romero-G{\'o}mez} {et~al.}(2007){Romero-G{\'o}mez}, {Athanassoula},
  {Masdemont}, \& {Garc{\'\i}a-G{\'o}mez}}]{RomeroGomez2007}
{Romero-G{\'o}mez}, M., {Athanassoula}, E., {Masdemont}, J.~J., \&
  {Garc{\'\i}a-G{\'o}mez}, C. 2007, \aap, 472, 63

\bibitem[{{Struck-Marcell}(1990)}]{StruckMarcell1990}
{Struck-Marcell}, C. 1990, \aj, 99, 71

\bibitem[{{Theys} \& {Spiegel}(1977)}]{TheysSpiegel1977}
{Theys}, J.~C. \& {Spiegel}, E.~A. 1977, \apj, 212, 616

\bibitem[{{Wetzel} {et~al.}(2013){Wetzel}, {Tinker}, {Conroy}, \& {van den
  Bosch}}]{Wetzel2013}
{Wetzel}, A.~R., {Tinker}, J.~L., {Conroy}, C., \& {van den Bosch}, F.~C. 2013,
  \mnras, 432, 336

\bibitem[{{Willett} {et~al.}(2013){Willett}, {Lintott}, {Bamford}, {Masters},
  {Simmons}, {Casteels}, {Edmondson}, {Fortson}, {Kaviraj}, {Keel}, {Melvin},
  {Nichol}, {Raddick}, {Schawinski}, {Simpson}, {Skibba}, {Smith}, \&
  {Thomas}}]{Willett2013}
{Willett}, K.~W., {Lintott}, C.~J., {Bamford}, S.~P., {et~al.} 2013, \mnras,
  435, 2835

\bibitem[{{Willmer}(2018)}]{willmer2018}
{Willmer}, C. N.~A. 2018, \apjs, 236, 47

\bibitem[{{Wong} {et~al.}(2006){Wong}, {Meurer}, {Bekki}, {Hanish},
  {Kennicutt}, {Bland-Hawthorn}, {Ryan-Weber}, {Koribalski}, {Kilborn},
  {Putman}, {Heiner}, {Webster}, {Allen}, {Dopita}, {Doyle}, {Drinkwater},
  {Ferguson}, {Freeman}, {Heckman}, {Hoopes}, {Knezek}, {Meyer}, {Oey},
  {Seibert}, {Smith}, {Staveley-Smith}, {Thilker}, {Werk}, \&
  {Zwaan}}]{Wong2006}
{Wong}, O.~I., {Meurer}, G.~R., {Bekki}, K., {et~al.} 2006, \mnras, 370, 1607

\bibitem[{{Wu} \& {Jiang}(2012)}]{WuJiang2012}
{Wu}, Y.-T. \& {Jiang}, I.-G. 2012, \apj, 745, 105

\bibitem[{{Wu} \& {Jiang}(2015)}]{Wu2015}
{Wu}, Y.-T. \& {Jiang}, I.-G. 2015, \apj, 805, 32

\bibitem[{{Zou} {et~al.}(2019){Zou}, {Zhou}, {Fan}, {Zhang}, {Zhou}, {Peng},
  {Nie}, {Jiang}, {McGreer}, {Cai}, {Chen}, {Chen}, {Dey}, {Fan}, {Findlay},
  {Gao}, {Gu}, {Guo}, {He}, {Jiang}, {Jin}, {Kong}, {Lang}, {Lei}, {Lesser},
  {Li}, {Li}, {Lin}, {Ma}, {Maxwell}, {Meng}, {Myers}, {Ning}, {Schlegel},
  {Shao}, {Shi}, {Sun}, {Wang}, {Wang}, {Wang}, {Wei}, {Wu}, {Wu}, {Wu},
  {Yang}, {Yang}, {Yuan}, \& {Yue}}]{2019ApJS..245....4Z}
{Zou}, H., {Zhou}, X., {Fan}, X., {et~al.} 2019, \apjs, 245, 4

\end{thebibliography}
%

\end{document}